# Spatial social networks identified from urban group travel

Huijun Sun[1], Kangli Zhu[1], Jianjun Wu[1,2*], Daqing Li[3*], Ziyou Gao[1], Haodong Yin[1], Yunchao Qu[1], Xin Yang[1], Hao Liu[4]

1. State Key Laboratory of Rail Traffic Control and Safety, Beijing Jiaotong University Beijing, 100044, China
2. Key Laboratory of Transport Industry of Big Data Application Technologies for Comprehensive Transport, Ministry of Transport, Beijing Jiaotong University, 100044, China
3. School of Reliability and Systems Engineering, Beihang University, Beijing 100191, China
4. Beijing Transportation Information Center, Beijing, 100161, China

**While the individual travel implicates the trace of individual mobility decision, group travels signify the possible social relationship behind these traces. Different from online social network, spatial interaction between individuals is a critical yet unknown dimension to understand the collective behaviors in a city. In this paper, based on over 127 million trips in Beijing metro network, we develop a method to distinguish the group travel of friends from the encounter travel of familiar strangers. We find travels of friends are among the most predictable groups. These identified friendships are interwoven and form a friendship network, with structural properties different from encounter network. The topological role of individuals in this network is found strongly correlated with her travel predictability. The overall time savings of about 34190 minutes after redistribution of inefficient group traveler with flexible travel purposes shows the potential of designing specific traffic fares for group travel. Our identification and understanding of group travel may help to develop and organize new traffic mode in the future smart transportation.**

Human travel in a city is essential to understand decision-making process in other urban activities, especially in large cities[1]. Emerging research of human mobility patterns in the city can influence and even alter urban planning, traffic forecasting, business activity modelling, and emergence management[2-4]. The past decade has witnessed an increasing interest in the discovery of individual mobility through data analysis as well as theoretical modelling[5-6]. Instead of individual mobility, group travel, as the most common mode in the public transportation, is critical for predicting and managing colletive decisions due to peer effect[7]. Specifically, for sharing transportation[8], smart traffic and other future traffic modes, learning the group travel in a city will become the prerequisite for design and optimization of the whole system.

Human mobility has been studied since the availability of different datasets both for unimodal and multimodal transportation[9-11]. Travel trajectories are found to follow the scaling law of levy flights[12, 13]. For example, the trip distance distribution is found to follow a general power-law distribution, based on the circulation data of bank notes in the United States[12]. While these studies mainly focus on the travel pattern across different cities, mobility inside the city has also been studied. Based on mobile phone data[14-16], most individuals are found to prefer traveling within a short radius, while others cover long distances on a regular basis[17]. Moreover, interesting travel pattern has also been found that most people will revisit some locations repeatedly and frequently, while others explore more places of large distances[18, 19].

No man is an island. The studies above mainly focus on the travel pattern or mobility of a single person, while as a common mode for public transportation, group travel has rarely been touched. Encounters, know as familiar strangers proposed by Stanley Milgram[20], are found with the temporal regularity and the rhythmic interactions for the encounters, based on the travel smart card data in Singapore bus system[21]. Familiar strangers refer to a social phenomenon in a city where people are observed to be together repeatedly over a certain amount of time without any direct interactions. Actually, the deep interaction between individuals during their travel does not exist in encounter strangers, which only appear in group travel of friends with strong social ties. These spatial social relationship will influence and determine the decision-making process in the collective travels as well as other activities, such as repurchase intention of tourists[22-24] and group hunting[25, 26].

While knowing the group behavior with strong social tie is critical, identifying these groups from the noisy travel records is challenging. In daily life, people may choose to travel with their families, friends or colleagues on intention for school, working, shopping, touring, etc. Meanwhile, identification of hidden social ties from human movements in a larger spatial and temporal scope requires large datasets and corresponding methods, rather than the visible relation between individuals in online social networks. In this paper, based on a huge amount of public traffic data, we propose a method to classify group travels and analyze the urban mobility features of different groups. We find travels of friends are among the most predictable groups. Furthermore, these friendships are interwoven and form a friendship network with power law degree distributions. Predictability of single person depends on the degree of this node in the friendship network identified from group travel.

**Results**

**Classifying group travel.** Our analysis of group travel patterns is based on over 127 million trips in Beijing metro network from March 1st to March 31st, 2014. (See Methods for details). Group travel here is defined as a couple of passengers who enter and leave the identical starting and ending station, within one-minute time difference (Figure 1a). Then we can count the total times of these two individuals for group travel in a given month, which can cover a set of different location pairs (as shown in Fig. 1b). We develop a phase-transition-method to finding critical threshold (see Methods), where the finding of critical $\tau_c$ is presented in figure1c-f.

In Fig.1c-d, the statistics of group network including the number of components and largest component size are explored. After a phase transition point of 6, the largest component S becomes stable. Then, by studying the degree distribution of potential group networks under different threshold, we find that they follow the power law as shown in Fig. 2e. Clearly, the power curves mostly overlap with each other when $\tau > 6$. Moreover, having investigated the spatial distribution of group travels, we found that patterns of home to work (HW) and work to home (WH) cover the majority of group travels as shown in Fig. 2d. Clearly, when $\tau = 6$, the proportion of HW and WH is lowest which indicates the diversity of friend travels. The results show that this critical value can include most of the repeated group travels when the cardholders travel together on purpose, without most encounters who run into each other with the same schedule (Figure 1b for an illustration of friends travelling together repeatedly).

Besides, trips with at least one group partner can be marked as group trips. For comparison, the familiar strangers are defined as two passengers entered the same metro stations within one minute more than 6 times and did not know each other (see S3 for the details of extracting encounter trips). The implication behind this is that passengers run into each other mainly on the platforms when they wait for the trains. Using the proposed group identification approach, 7,220,367 group trips are extracted from the AFC data of March, 2014, accounting for 5.698% of total trips, indicating the significance of a detailed investigation of these groups.

**Entropy of group travel.** Friends of group travel differentiate familiar strangers proposed by Stanley Milgram. As shown in Fig. 2a, the fluctuation of the number of hourly stranger trips is roughly the same as that of the total demand with a morning peak and afternoon peak on weekdays. This suggests that the group travels have similar choice for travel time as other cases, and many encounters are coincidences due to regular schedules on commuting hours. On the contrary, when we check the visiting pattern and the pattern similarity of spatial trajectories shown in Fig. 2b and 2c, group travels of friends are found distinct from the other types. These friend groups visit 1-2 locations with high frequency, while other types usually have more locations selected as trip destinations. We also investigate the travel entropy in Fig. 2d-2l. Compared with the strangers and the encounters, the heterogeneous entropy of friends have the strongest similarity among three types, suggesting that friends always have the same point of interests, trip preference or travel purpose.

**Social network identified from group travel.** The person that we choose to talk over the Internet is different from the person whom we travel together with. Traveling together on purpose shows far more strong social ties [27-29], at a larger cost of time and money, rather than word of mouth in online social networks. Fig. 3a displays some randomly selected components (upper panel of Fig. 3a) and the largest component (lower panel of Fig. 3a) of the aggregated group networks. As shown in Fig. 3a, travelers can form a social network composed of clusters of different sizes. It is shown in Fig. 3b that the cluster size follows a power law distribution: most clusters are small including few individuals, while there exist a few groups containing many individuals connected with their friends. This is also confirmed in degree distribution of Fig. 3c that some individuals have many friends sharing the trip choice, acting as hubs in the spatial social networks. Clearly, the spatial social network is constrained by the time cost in daily trips, since each trip cost a significant amount of time. Spatial social networks of group travel are distinct from encounter networks, which have more hubs yet weak interconnections. This generates a network with larger scaling exponent (2.93) than many online social networks [7, 30-32]. Interestingly, individuals in group travel are found to have preference for a given friend, as large link weight shown in Fig. 3d.

At the microscopic level, the spatial social networks are composed of various motifs. As shown in Fig. 3e, group travel of 2 persons is the most common type. The appearance of group travel is decreasing with the number of persons involved. For the groups with the same size, the appearance seems independent of group topology. For the group with 4 or 5 nodes, nodes in the group are mostly mutually connected, which reflects the strong community in the group network. This reflects the positive correlation between the edges and the distance in the network—a larger probability of being friends with the friends of friends.

**Entropy and connection.** Individual behaviors can be inferred from their friends. Given the identified spatial social network, we wish to know the entropy as predictability for individuals as nodes in social networks (see Methods). The entropy is found to have heterogeneity distribution in the whole spatial social network, suggesting the importance of finding the influential individuals. As shown in Fig. 4a-c, entropy is decreasing with group size. This suggests that when one individual is embedded in a larger social group, her behavior is more predictable. Furthermore, as shown in Fig. 4d-f, individual with more friends (large degree in spatial social network) identified from group travel is also more predictable. These findings enable us to understand and predict the individual choices within the embedded spatial social network, including traffic, evacuation and ride-sharing.

**The potential of group fare discount to improve efficiency.** For trips with shorter distances and longer trip duration using metro system, the inefficiency of metro system may catalyze the passengers to transfer mode of transportation, especially group travelers. This may improve the efficiency of public resources and decreases the occupation of the subway thus improve the level of service. An index is defined to capture the metro efficiency as $ddr = \frac{distance}{duration}$, which is the average speed of each trip. A lower value of $ddr$ implies inefficiency of using metro compared with other modes. In Fig. 5a, low $ddr$ OD pair implies more interchange times which make a long travel duration or geographical vicinity which magnify the proportion of waiting time in travel duration thus reduce the overall speed. In some sense, the synthesis of $ddr$ of a origin is a substitute for accessibility. Fig. 5b provide the spatial distribution of all the least 10% $ddr$ OD pairs. Fig. 5c-d provides the spatial distribution of passengers on the sections using the passengers between least 50% and 10% $ddr$ OD pairs. If guided to other mode of transportation, these links can experience a better level of service.

A group ticket policy can be made to redistribute the group travels temporally to improve the overall performance of the metro system. The flexible (non-commuting), inefficient (least $ddr$) passengers are redistributed temporally and a simulation is conducted using the passenger flows to evaluate the potential gains of this kind of policy. Fig. 5e-f gives the spatial distribution of section load and the reduced section load after temporal passenger redistribution of morning peak hour from 7:00 to 8:00.

The results are obtained from the mesoscopic passenger simulation algorithm developed by Yin et al.[34], which has been applied in analyzing passenger flow of Beijing subway. The results show the best reduced section load after passenger redistribution temporally is about 0.02. And the overall time savings is about 34190 minutes.

**Discussion**

In traffic modeling and engineering, travel behavior analysis plays an essential role, where current researchers treat each individual separately. Meanwhile, different from the travel patterns of independent passengers, our findings suggest that members in a group of travelers involve in interpersonal interactions frequently and possess a different travel choice logistic. They behave differently from member who travels alone, mainly due to coordinating each agent's schedule and preferences. Existing studies about group behavior mainly focus on the modeling of intra-household interactions since family members share various household resources and their trips can be complementary. Group travel can help to better model other collective traffic behaviors from the location of entertainment and leisure activities selection to departure time selection. Moreover, the temporal elasticity and relatively low efficiency of group travel shows the potential of a fare discount policy for groups to improve the overall performance of inner city transit system.

Bundled travel behavior is particularly useful for the emergency management in the traffic system. Since it takes time to travel together and the group travelers always travel to conduct a certain activity, the study of group travel behavior can assist the emergency management. When emergency occurs either at a subway station, bus stop during the travel or at a shopping mall, place of interest during the activity, those travelers with group partners may take their partners into consideration when choosing the evacuation or even escape path. This factor affects the evacuation efficiency, which is the main parameter for grouping walking modeling[26] and further for evacuation design and management.

Until now discovery of human patterns in the city is incomplete. Here, we develop a method to find friends in the large real travel data of urban rail transit AFC smart card has been developed. The patterns of group travel are analyzed compared with the others, strangers and encounters (familiar strangers). Based on a phase-transition method, 6 times of group travel in the metro network is found critical to distinct group from other travel types. The spatial temporal distribution of the group travel demand as well as the group ratio depicts a busy downtown frozen world and a peaceful suburban smiley world. Besides, the group travels help to reveal the distinction of a spatially constrained social network in the real world rather than the commonly studied online social network.

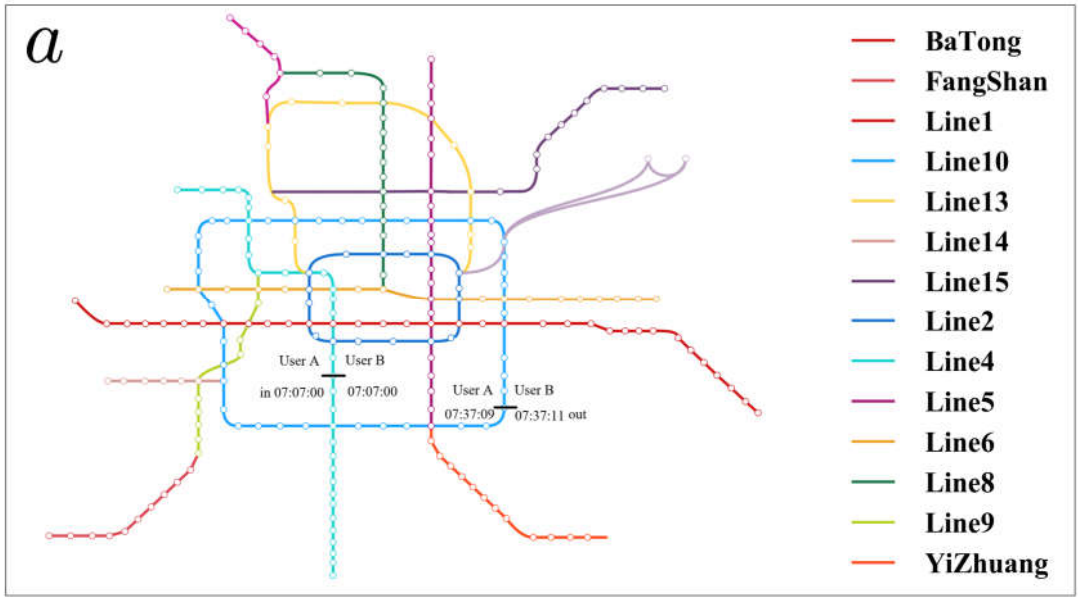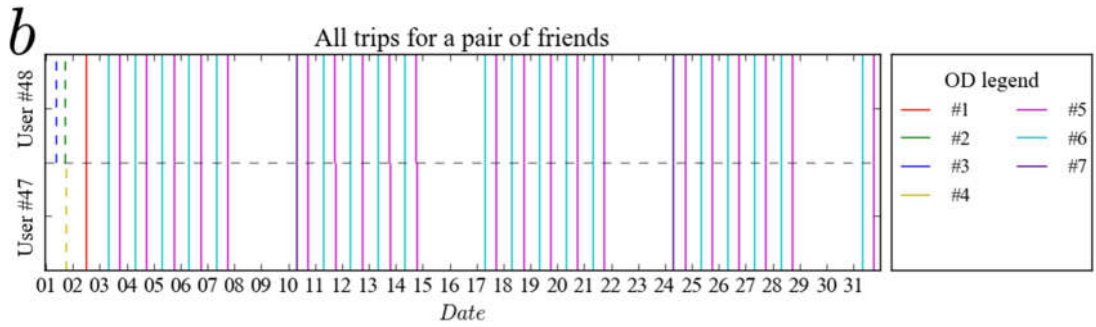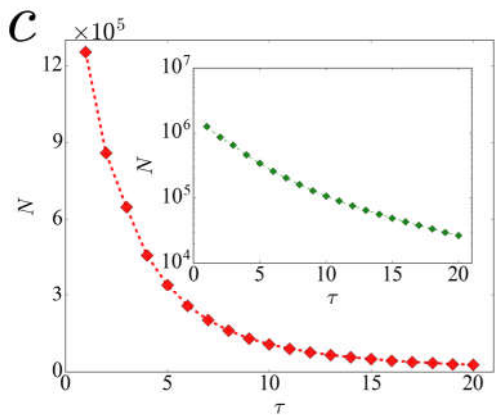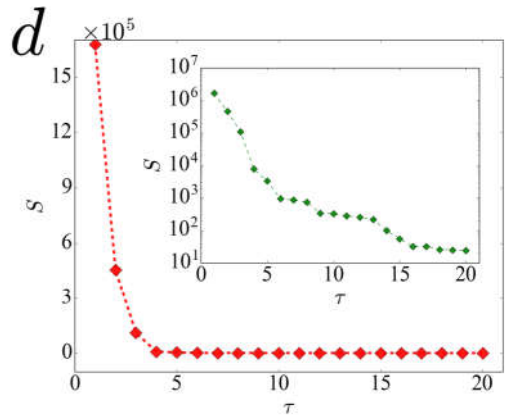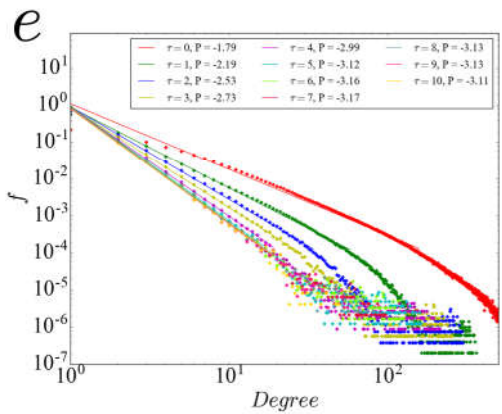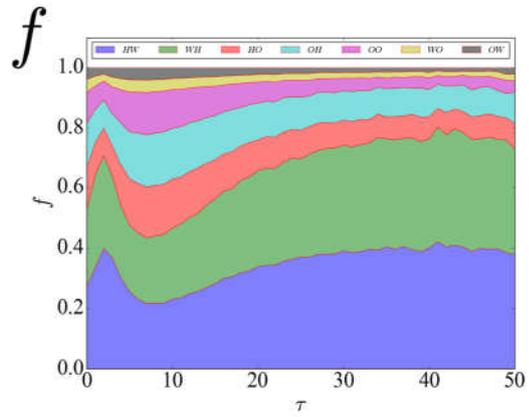

**Figure 1. The demonstration of group travels for a pair of travelers (friends) and results of the phase-transition-method to find the critical times and identify the group travel.** (a) Potential group trips enter and exit the same location within one minute. b) Potential friends are those who conducted potential group trips for more than a certain threshold times within one minute with the same entering and exiting stations. All the trips for two users over 31 days are shown. The lines above and below the horizontal dashed lines represent two users' trip respectively (lower ones for #47 and upper ones for #48). The vertical solid lines denote the trips #47 and #48 enter and exit the same station(different colors denote different OD pair) within one minute, while the vertical dashed ones were drawn when a user travel alone. In the temporal diagram of two passengers, the spatial trajectory of user#47 and user #48 coincides with each during the most of trips and stays with occasional differences, which is shown in the figure that the distance of most upper lines and lower lines with the same color are very close. c) The number of components under different thresholds. d) The size of the largest components under different thresholds. The largest component S becomes stable when $\tau>6$ .e) The degree distribution of potential group networks under different threshold. The power firstly increases and then decreases with the power law curve mostly overlapping with each other when $\tau > 6$. Besides, an interesting finding is that the power exponent is about 3 agreeing with the results of most growing scale-free network with preference attachment in the real world. f) The origin and destination attributes proportion for different $\tau$. The most and second frequently visited locations are regarded as home(H) and work(W) place. Together with other places(O), the trips are classified into HW/WH/HO/OH/WO/OW/ according to origins and destinations. The proportion of commute trips are lowest when threshold is 6.

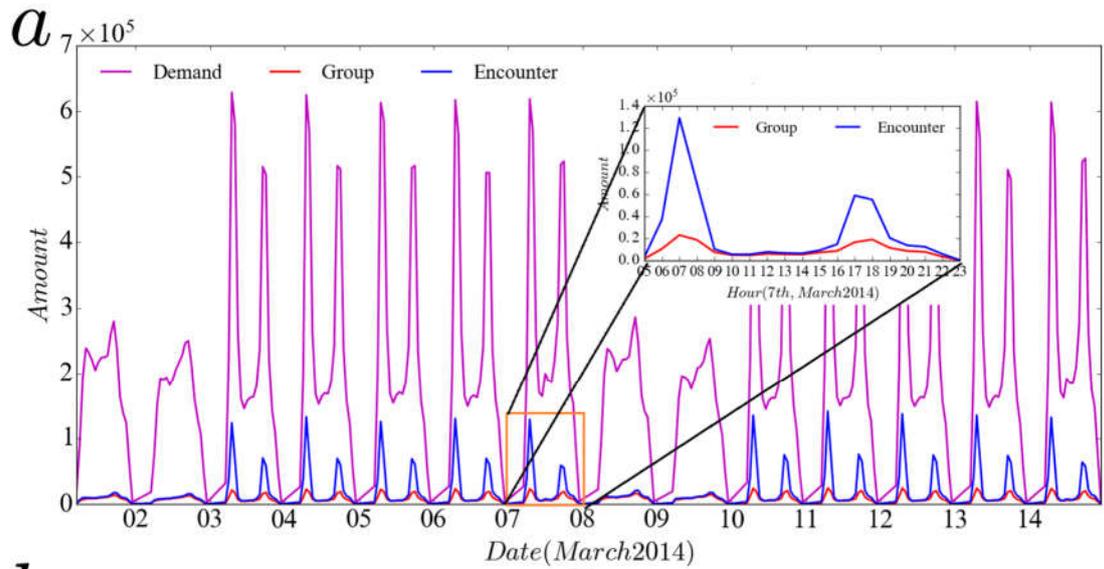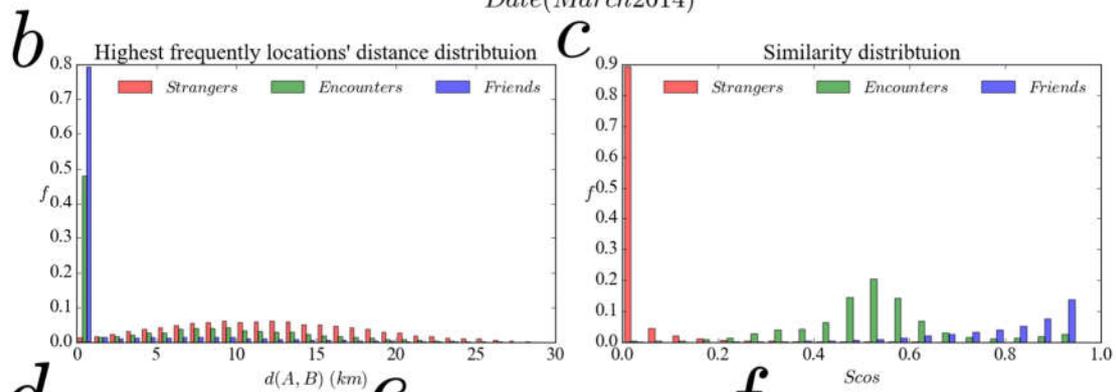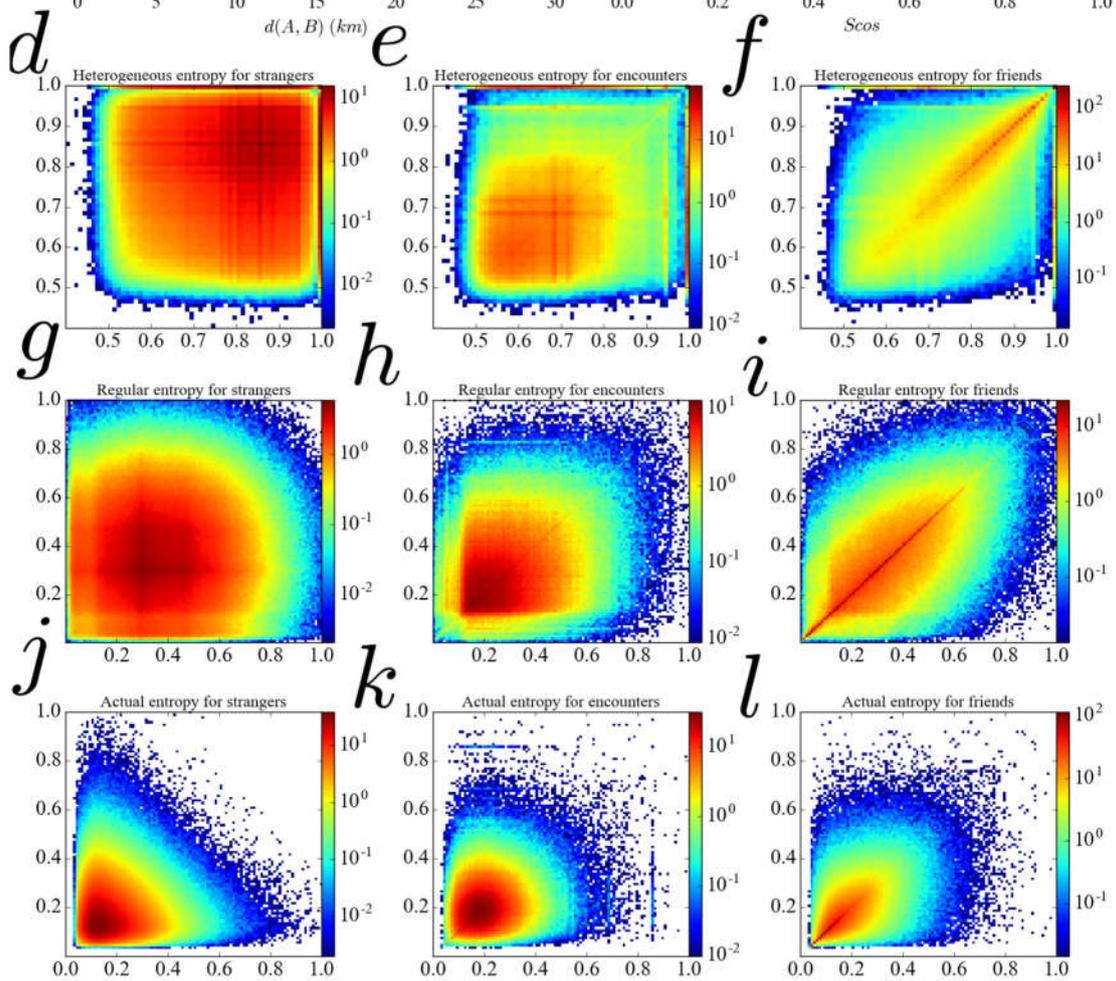

**Figure 2. The spatial temporal distribution of group travels.** a) The number of accumulative strangers, encounters and friends in the collected data. Stranger trips make up the most of total demand, group trips with friend nearly accounts for all the encounter trips conducted by familiar strangers on off-peak hours while most coincidences happen at weekday rush hours. The inset show the number of encounter and group trips on March $7^{th}$. b) The displacements distributions of strangers, encounters and friends respectively. Clearly, the distance of strangers follows normal distribution, and encounters shows a inclination of distance zero. However, for friends, the distribution $p(d(A, B))$ exhibits a much smaller distribution: showing different travel behavior compared with the "familiar strangers" or encounters. c) The probability distribution of spatial cosine similarity $scos(A, B)$. Obviously, the strangers display diversified travel experience because the similarity distribution $scos(A, B)$ is closed to zero. However, for encounters and friends, the similarity distribution for two passengers is significantly more than zero, especially for the latter. d)-f) Heterogeneous entropy comparison for strangers, encounters and friends. g)-i) Regular entropy comparison for strangers, encounters and friends. Personal regular entropy over the time slots can be used to explain the regularity of an individual. The correlation of the regular entropy of different kind of users are friends > encounters > strangers. j)-l) Actual entropy comparison for strangers, encounters and friends. Actual entropy denotes the predictability of a user. The correlation of the actual entropy of different kind of users are friends > encounters > strangers.

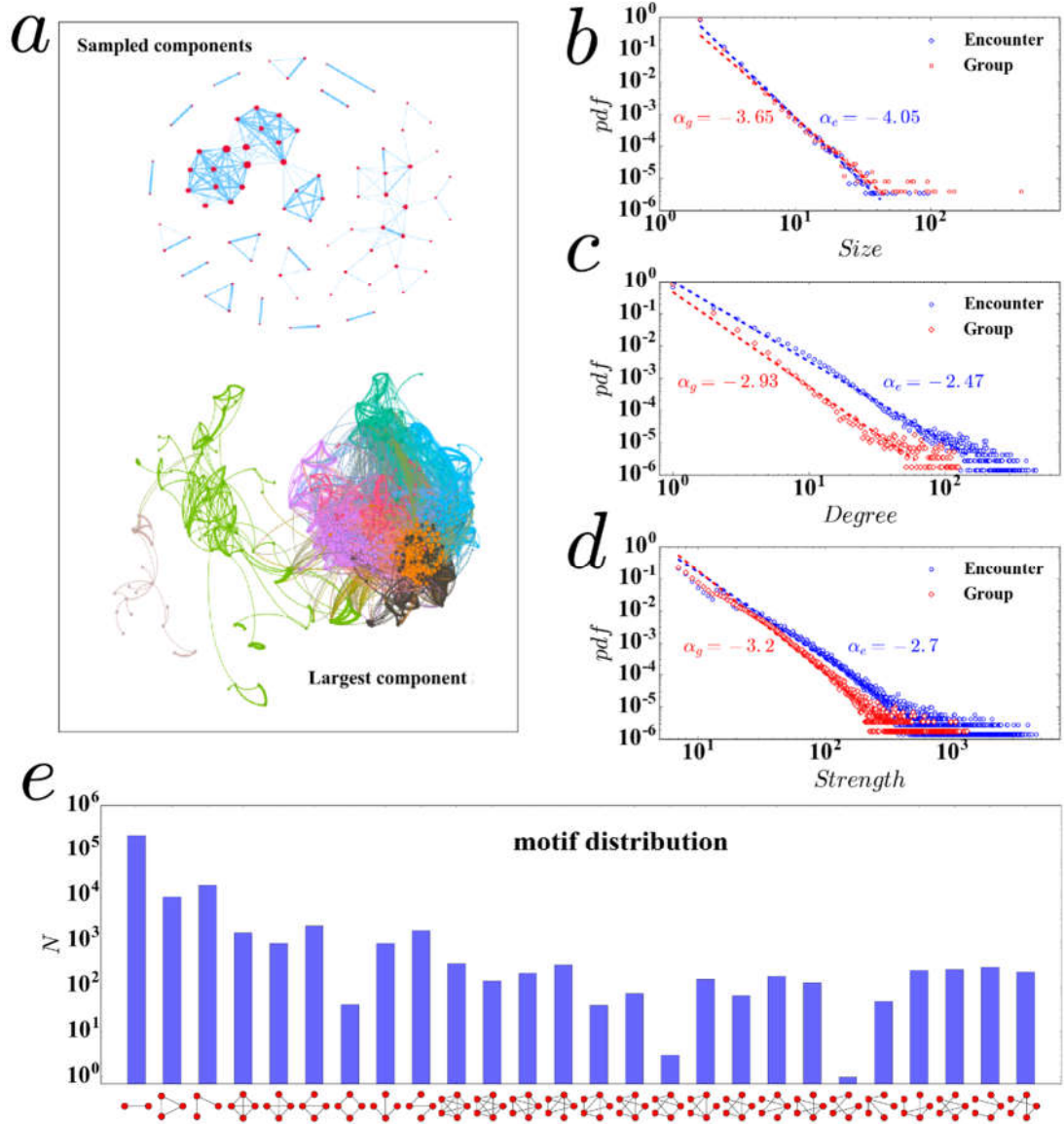

**Figure 3. Characteristics of group network.** a) Sampled connected components. The upper one are randomly sampled components and the size of each node indicates degree in unweighted network, i.e., the number of people one have been traveled in group with at least six times during a month. The lower one displays the maximal connected component. Size of each node indicates degree in unweighted network, i.e., the number of people one have been traveled in group with at least six times during a month; color of each edge shows the community it belongs to using Gephi. Besides, thickness of each edge indicates the weight, i.e. the number of time that two individual have travelled in group. b)–d) Degree distribution, strength distribution and Size distribution of connected component. The ones for encounter networks is also shown for comparison. e) Static group motif distribution. The horizontal axis is connected sub-graph structure with the sum of node degree increasing from left to right; the vertical axis is the number of connected sub-graphs corresponding to the structure (See S4 for details).

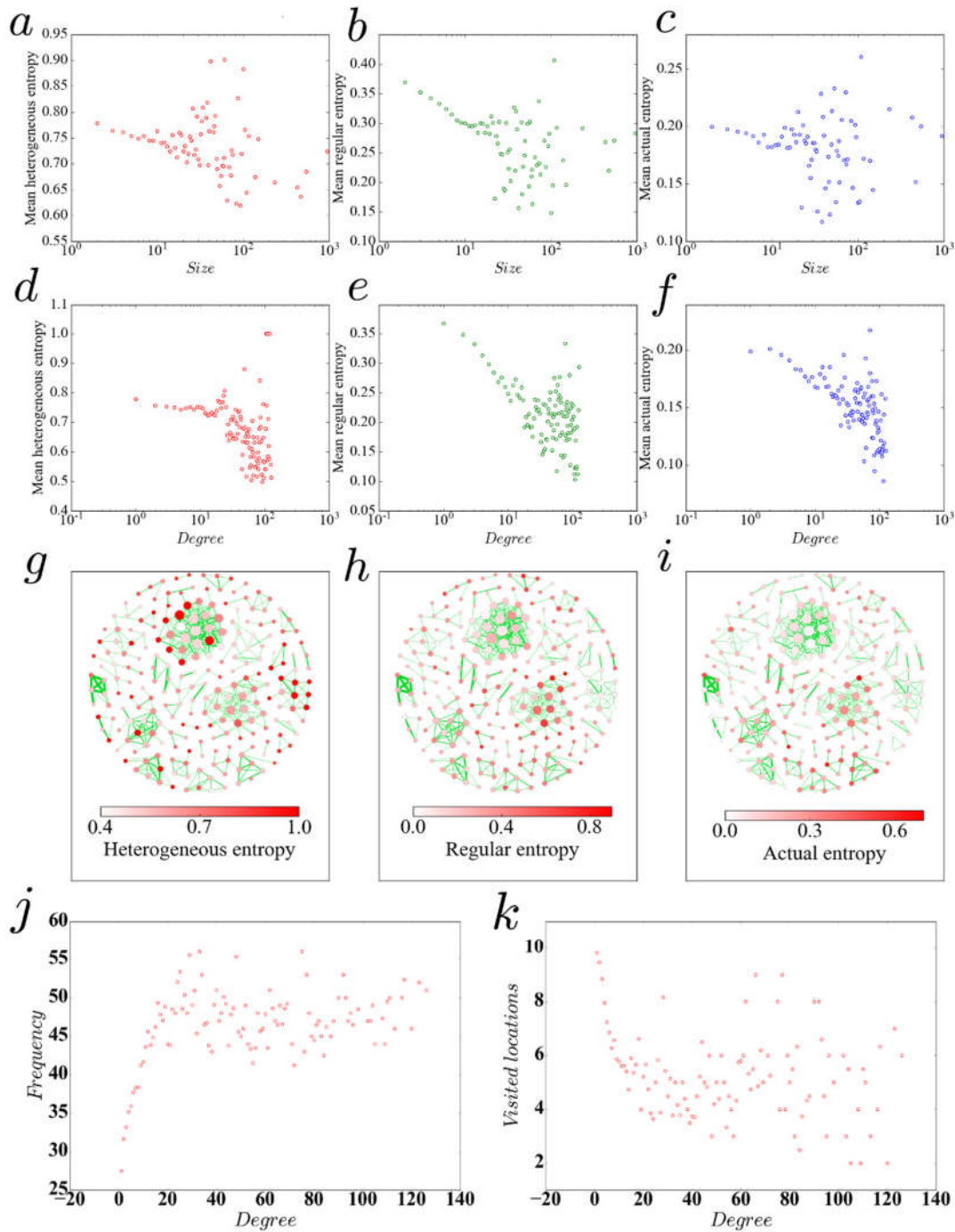

**Figure 4. Entropy distribution in the social group networks.** a)-c) The mean heterogeneous, regular and actual entropy of users in different component size. For little components, the entropies in a larger component are larger than those in a smaller one. d)-f) The mean heterogeneous, regular and actual entropy change with degree. For nodes with less edges, the entropies are larger than those with more edges. g)-i) The heterogeneous, regular and actual entropy of sampled components, where nodes correspond to users, the color of the nodes are proportional to the three entropies according to the color bars below and the size of nodes corresponds to the degree of users. j)-k) Trip frquency and visited locations of a user versus its degree in the group network. The more friends a user have, the more trips he has conducted during the month and the less locations he has visited. This conclusion is consistent with the entropy.

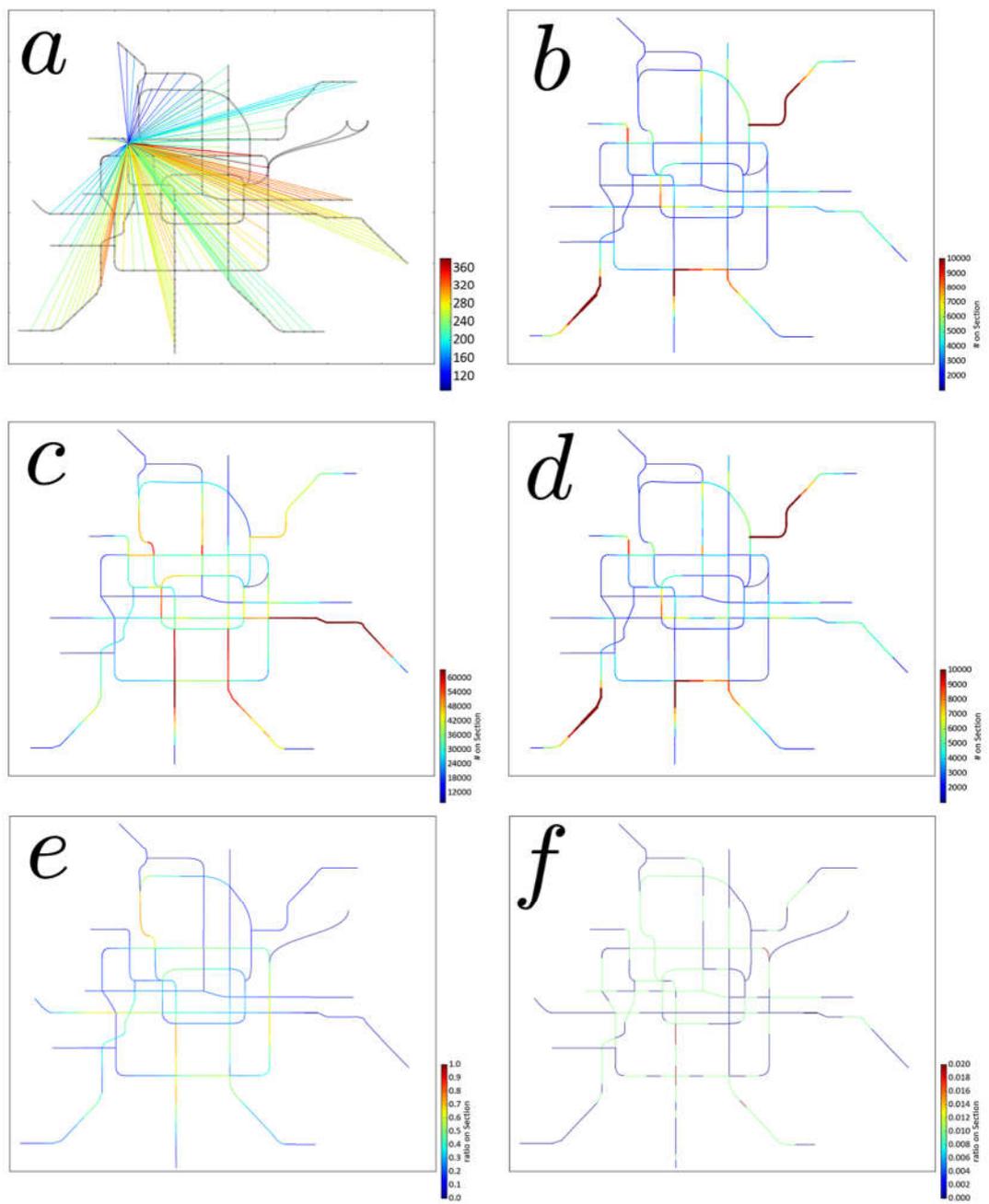

**Figure 5. The illustration of spatial distribution of low *ddr* OD pairs** a) An example of *ddr* distribution using OD pairs originated from one station b) The spatial distribution of all the least 10% *ddr* OD pairs. c) The spatial distribution of passengers on the sections using the passengers between least 50% *ddr* OD pairs. d) The spatial distribution of passengers on the sections using the passengers between least 10% *ddr* OD pairs e) the spatial distribution of section load. f) The spatial distribution of reduced section load after passenger redistribution temporally.